\documentstyle[12pt,epsfig]{article}                                
\begin{document}

\titlepage

\begin{large}
\begin{bf}
 PT-symmetry an  ordered  quantum system 

\end{bf}

 \vspace{0.1cm}

 Biswanath Rath

\end{large}

\vspace{0.1cm}

Department of Physics,
 Maharaja Sriram Chandra Bhanj Deo University,
 Takatpur, Baripada -757003, Odisha, INDIA

(*biswanathrath10@gmail.com)

\begin{bf}
Abstract
\end{bf}

We argue by saying that due to conservation of  energy($<H>_{n} \rightleftharpoons <K.E>_{n} + <P.E>_{n}$)  PT-symmetry Hamiltonian $H=p^{2} - (ix)^{N} $ is a highly ordered system.
Further, it is found that $<K.E>_{n}  \gg   <P.E>_{n} $.

\vspace{0.1cm}

\begin{bf}
PACS no-
\end{bf}

 03.65.Ge

\vspace{0.1cm}

\begin{bf}
Key words:
\end{bf}
PT-symmetry , negative potential 

\vspace{0.1cm}

\vspace{0.1cm}

\begin{bf}
1.Introduction
\end{bf}

Energy conservation is an universal principle. Physically it means that 

\vspace{0.1cm}

\begin{bf}

Kinetic energy + Potential energy = Total energy = Constant 

\end{bf}

\vspace{0.1cm}

This universal principle remains valid in classical as well as quantum mechanics

of real or complex systems. However, in quantum mechanics  above relation should  
be read as 

\begin{equation}
<K.E>_{n} + <P.E>_{n} = <E>_{n} 
\end{equation}

 Hence, it is believed that energy  distibution 
is possible in all measurable systems[1]. In this context, we find there has
 been considerable interest in complex systems involving unbroken PT-symmetry. 
Mathematically[2] 

\begin{equation}
[H,PT]=0
\end{equation}

Here, P stands for parity operator having space reflection property
\begin{equation}
P x P^{-1}= -x
\end{equation}
\begin{equation}
P p P^{-1}= -p
\end{equation}
Similarly T-stands for time reversal property 
\begin{equation}
T i T^{-1}= -i
\end{equation}
\begin{equation}
T p T{-1}= -p
\end{equation}
\begin{equation}
T x T^{-1}= x
\end{equation}
 Under this transformation , the commutation relation remains invariant [3]

\begin{equation}
[x,p] = 0
\end{equation}

In fact an interesting result is that  the   PT-symmetry deals with negative potentials,whose corresponding Hamiltonian is 
\begin{equation}
H= p^{2}-(ix)^{N} \rightarrow N\gg 2
\end{equation}
whose energy eigenvalue is of positive in nature 
\begin{equation}
E_{n} \sim \Biggl[\frac{\Gamma(3/2+1/N)\sqrt{\pi}(n+1/2)}{\sin(\pi/N)\Gamma(1+1/N)}\Biggr]^{2N/N+2} \hspace{1.0in} ( n \rightarrow  \infty)
\end{equation}
The above formula is derived using WKB method[2].As stated above, WKB approach 
gives good estimates of energy eigenvalue for largeer value of  n.

 Further Bender,Berry,Meisinger,Savage and Simsek[4]solved the WKB method
 for the same potential using the formula
\begin{equation}
(n+\frac{1}{2})\pi = \int_{x_{-}}^{x_{+}} dx \sqrt{(E + (ix)^{N}}
\end{equation}
 and reported that 
\begin{equation}
E_{n} \sim \Biggl[\frac{\Gamma(3/2+1/N)\sqrt{\pi}(n+1/2)}{\sin(\pi/2N)\Gamma((1+N)/2N)}\Biggr]^{2N/N+2} \hspace{1.0in} ( n \rightarrow  \infty)
\end{equation}
Further it has been reported that above two different formulas are valid only 
for $N\gg 2 $. However we find for N=6 above two formulae are not suitable because the term $-(ix)^{N}$ is just a sextic operator with positive sign i.e$ (+x^{6})$.
Apart from the above, we find another formula as[5] 
\begin{equation}
E_{n} \sim \Biggl[\frac{\Gamma(3/2+1/N)\sqrt{\pi}(n+1/2)}{\cos(-\frac{\pi}{2}+
\frac{\pi}{N})\Gamma(1+1/N)}\Biggr]^{2N/N+2} \hspace{1.0in} ( n \rightarrow  \infty)
\end{equation}
In above the original formula has been modified considering the potential$-V(x)=
(ix)^{N}$.
In this context, we notice Ahmed,Bender and Berry[6] reported another  
  general non-polynomial negative PT-symmetry operator having Hamiltonian  
\begin{equation}
H=p^{2} -  x^{2K+2} \rightarrow K=1,2,3.......
\end{equation}
whose energy eigenvalue using WKB approximation reads as 

\begin{equation}
E_{n}\sim \Biggl[\frac{(n+1/2)\sqrt{\pi}(K+2)\Gamma{[(K+2)/(2K+2)]}}{\Gamma{[1/(2K+2)]}
\cos{[\pi/(2K+2)]}}\Biggr]^{(2K+2)/(K+2)}
\end{equation}

Now we have four different formulae on $V(x)=-x^{4}$. All the formulae bring out the unique feature saying that inverted negative quartic potential has only positive discrete energy levels. However, which one the reader will use to
 verify to know the $<K.E>_{n}$ ? . This answer can hardly be answered at present . Hence, the only possibility is numerical calculation. Below we consider few model PT-symmetry potentials.

\begin{bf}
2. Model PT-symmetry complex potentials
\end{bf}

Here we consider few complex potentials as 

\begin{equation}
H_{1}= p^{2} + m^{2} - (ix)^{N}  \hspace{1.in}  (N=4;m^{2}=1)
\end{equation}

\begin{equation}
H_{2}= p^{2} - x^{4} - 2 i x  
\end{equation}

\begin{equation}
H_{3}= p^{2} + i x^{3} 
\end{equation}

\begin{equation}
H_{4}= p^{2} + i x^{5} 
\end{equation}

\begin{equation}
H_{5}= p^{2} + i x^{7} 
\end{equation}

\begin{equation}
H_{6}= p^{2} - x^{6}
\end{equation}

\begin{equation}
H_{7}= p^{2} - x^{8}
\end{equation}
 and 
\begin{equation}
H_{8}= p^{2} + \frac{1}{4} x^{2} - x^{4}
\end{equation}
\begin{equation}
H_{9}= p^{2} +  x^{4} + 2ix    \hspace{1.0in} [11]           
\end{equation}
\begin{equation}
H_{10}= p^{2} + \frac{x^{4}}{1+x^{2}} + ix
\end{equation}

\begin{bf}
3. Method 
\end{bf}

 Here we use matrix diagonalisation method(MDM) on solving the eigenvalue relation[10,11].

\begin{equation}
H|\Psi> = E|\Psi>
\end{equation}
with
\begin{equation}
|\Psi>=\sum_{m} A_{m}|m>
\end{equation}
where $|m>$ satisfies the relation
\begin{equation}
[H_{0}=p^{2}+x^{2}]|m>=(2m+1)|m>
\end{equation}

\begin{bf}
4.Results and Discussion
\end{bf}

In this papaer, we correctly compute both $<K.E>_{n}$ and $<P.E>_{n}$ and
 reflect the same in tables.1-7. For $V(x)=ix^{5}$ and $V(x)=ix^{7}$, the corresponding potential energy contribution is zero. This motivates to present typical calculations involving $H_{8}$ ,where we also find the potential energy contribution. In fact we find it also very near to zero. However interested readers can extend this calculation to suitable value of $m^{2}$. One essential thing we believe is that like hermiticity , all physical quantities in PT-potentials are 
measurable. So we feel unbroken PT-symmetry models are highly ordered system.

\pagebreak

\begin{table}

 Table-1: Energy conservation in $H_{1} = p^{2} + x^{2} - x^{4}$

\vspace{1.0cm}
\begin{tabular}{ c c c c c } \hline
level & $<p^{2}> $  & $ <x^{2}-x^{4}> $& Total & Correct  \\ \hline
0  & 1.747 7 6 & -0.595 4 & 1.152 3 & 1.152 2\\
1 &3.881 7  &  1.217 2 & 5.098 9  & 5.098 9 \\
2 & 6.898 9 & 3.541 7 &10.440 6 & 10.440 6 \\
3 & 12.112 0 &4.589 6 &16.701 6 & 16.701 5\\  \hline
\end{tabular}

\end{table}

\begin{table}

 Table-2: Energy conservation in $H_{2} = p^{2} - x^{4} - 2ix $

\vspace{1.0cm}
\begin{tabular}{ c c c c c } \hline
level & $<p^{2}> $  & $ <-x^{4} - 2ix> $& Energy & Previous[7]  \\ \hline
0  & 1.687 5  & -1.687 5 &0 &0 \\
1  & 3.576 1  &  -0.178 0& 3.398 2& 3.398 2\\
2  &6.963 8 & 1.736 7& 8.700 5 & 8.700 5\\ 
3 & 12.7432 & 2.234 6 &14.977 8 & 14.977 8\\  \hline
\end{tabular}
\end{table}

\begin{table}

 Table-3: Energy conservation in $H_{3} = p^{2} + ix^{3}$

\vspace{1.0cm}
\begin{tabular}{ c c c c c } \hline
level & $<p^{2}> $  & $ <ix^{3}> $& Energy & Previous[2,12]  \\ \hline
0  & 1.246 7 & - 0.090 4 & 1.156 3 & 1.156 3\\
1  & 2.971 7 & 1.137 5 & 4.109 2& 4.109 2\\
2  & 4.808 9 & 2.713 4& 7.562 3& 7.562 3 \\ 
3 & 6.645 0 & 4.669 4 &11.314 4 & 11.314 3\\  \hline
\end{tabular}
\end{table}

\begin{table}

 Table-4: Energy conservation in $H_{4} = p^{2} + ix^{5}$

\vspace{1.0cm}
\begin{tabular}{ c c c c c } \hline
level & $<p^{2}> $  & $ <ix^{5}> $& Total & Correct \\ \hline
0  & 1.164 8 & 0 &1.164 8 & 1.164 8\\
1  & 4.363 8 & 0 & 4.363 8 &4.363 8\\
2  & 8.955 2 & 0 &8.955 2 & 8.955 2 \\ 
3 & 14.417 8 & 0 &14.417 8 &14.417 8\\  \hline
\end{tabular}
\end{table}

\begin{table}

 Table-5: Energy conservation in $H_{5} = p^{2} + ix^{7}$

\vspace{1.0cm}
\begin{tabular}{ c c c c c } \hline
level & $<p^{2}> $  & $ <ix^{7}> $& Total & Correct \\ \hline
0  &1.224 7  & 0 &1.224 7 & 1.224 7\\
1  &4.721 5 & 0 &4.721 5 &4.721 5\\
2  &10.075 4 & 0 &10.075 4 &10.075 4 \\ 
3 & 16.872 5 & 0 &16.872 5 &16.872 5\\  \hline
\end{tabular}
\end{table}

\begin{table}

 Table-6: Energy conservation in $H_{6} = p^{2} - x^{6}$

\vspace{1.0cm}
\begin{tabular}{ c c c c c } \hline
level & $<p^{2}> $  & $ <-x^{6}> $& Total & Correct \\ \hline
0  &2.002 7  &-0.647 8 &1.354 9 & 1.354 9\\
1  &6.047 5 &-0.784 9 &4.721 5 &4.721 5\\
2  &11.139 4 &0.095 5 &11.234 9 &11.235 0 \\ 
3 & 17.518 6 &0.979 3 &18.497 9 &18.497 9\\  \hline
\end{tabular}
\end{table}

\begin{table}

 Table-7: Energy conservation in $H_{7} = p^{2} - x^{8}$

\vspace{1.0cm}
\begin{tabular}{ c c c c c } \hline
level & $<p^{2}> $  & $ <-x^{8}> $& Total & Correct \\ \hline
0  &5.137 4  & -3.777 6 &1.359 9 & 1.359 9\\
1  &11.218 7 & - 5.898 2 &5.320 5 &5.320 5\\
2  &17.911 0 &-6.351 1 &11.559 9 &11.559 9 \\ 
3 & 26.187 7 &-6.534 9 &19.652 8 &19.652 8\\  \hline
\end{tabular}
\end{table}

\begin{table}

 Table-8: Energy conservation in $H_{8} = p^{2} + \frac{1}{4} x^{4} - x^{4}$

\vspace{1.0cm}
\begin{tabular}{ c c c c c } \hline
level & $<p^{2}> $  & $ <0.25 x^{2} - x^{4}> $& Energy & Correct  \\ \hline
0  &1.355 8 & 0.004 5 & 1.360 3 & 1.360 3\\ \hline
\end{tabular}
\end{table}

\begin{table}

 Table-9: Energy conservation in $H_{9} = p^{2} + x^{4} + 2ix$

\vspace{1.0cm}
\begin{tabular}{ c c c c c } \hline
level & $<p^{2}> $  & $ <x^{4}+ 2ix> $& Total & Correct [11]\\ \hline
0  &1.245 5  & 0.376 3&1.630 7 & 1.630 7 \\
1  &2.939 6 & 0.881 9 &3.821 5 &3.821 5      \\
2  &5.799 0 & 1.739 7 &7.538 7 &7.538 6 \\ 
3 & 9.007 3 & 2.702 2 &11.709 5&11.709 5\\  \hline
\end{tabular}
\end{table}

\begin{table}

 Table-10: Energy conservation in $H_{10} = p^{2} + \frac{x^{4}}{1+x^{2}}+ ix$

\vspace{1.0cm}
\begin{tabular}{ c c c c c } \hline
level & $<p^{2}> $  & $ <\frac{x^{4}}{(1+x^{2}} + ix> $& Total & Correct \\ \hline
0  & 0.772 2 &0.319 2 &1.094 4 & 1.094 4\\
1  &1.914 9 & 0.743 2 &2.658 1 &2.658 1 \\
2  &3.472 7 & 1.214 0 &4.686 7 &3.472 7 \\ 
3 & 4.993 2 & 1.584 7 &6.578 4 &6.578 5\\  \hline
\end{tabular}
\end{table}

\end{document}